\documentclass{aa}
\usepackage{graphicx,times}

\newcommand{\xmm}{{\it XMM-Newton}}
\newcommand{\chandra}{{\it Chandra}}
\newcommand{\rgs}{{RGS}}
\newcommand{\epic}{EPIC}
\newcommand{\mos}{EPIC-MOS}
\newcommand{\pn}{EPIC-pn}

\begin{document}

\title{Synoptic study of the SMC SNRs using \xmm}

\author{K.J. van der Heyden\inst{1},
 J.A.M. Bleeker\inst{1},
 J.S. Kaastra\inst{1}}
 
\offprints{K.J. van der Heyden}

\institute{SRON National Institute for Space Research, 
Sorbonnelaan 2, 
3584 CA Utrecht, The Netherlands
\\email: K.J.van.der.Heyden@sron.nl \\} 
 
\titlerunning{\xmm\ observations of the SMC SNRs} 
\authorrunning{K.J. van der Heyden et al.} 
 
\date{Received ; accepted } 
 
\abstract{

We present a detailed X-ray  spectral analysis of 13 supernova remnants (SNR) in 
the Small
Magellanic Cloud (SMC). We apply both single-temperature non-equilibrium
ionisation models and models based on the Sedov similarity solution, where
applicable. We also present detailed X-ray images of individual SNRs, which
reveal a range of different morphological features. Eight remnants,
viz DEM S\,32, IKT 2, HFPK 419, IKT 6, IKT 16, IKT 18 and IKT 23, are 
consistent
with being in their Sedov evolutionary phase. IKT 6 and IKT 23 both have a
clear shell like morphology with oxygen-rich X-ray emitting material in the
centre. We draw attention to similarities between these two remnants and the
well studied, oxygen-rich remnant IKT 22 (SNR 0102-72.3) and propose that they
are more evolved versions of IKT 22. IKT 4, IKT 5, DEM S\,128 and IKT 5 are
evolved remnants which are in, or in the process of entering, the radiative 
cooling
stage. We argue that the X-ray emission from these four remnants is most 
likely
from the ejecta remains of type Ia SNe. Our modeling allow us to derive
estimates for physical parameters, such as densities, ages, masses and initial
explosion energies. Our results indicate that the average SMC hydrogen density
is a factor of $\sim 6$ lower as compared to the Large Magellanic Cloud. 
This
has obvious implications for the evolution and luminosities of the SMC SNRs. We 
also
estimate the average SMC gas phase abundances for the elements O, Ne, Mg, Si 
and
Fe.

\keywords{ISM: shock waves, nucleosynthesis, abundances, supernova remnants -- 
ISM: --X-rays} 
}

\maketitle

\section{Introduction} 

The low interstellar absorption and relative closeness of the Large and Small
Magelanic Clouds (LMC and SMC) allows for the study of individual X-ray sources 
in these
galaxies. For supernova remnants (SNRs) in particular, the low absorption makes
it possible to detect X-rays in the important 0.5--3.0 keV energy band, which
includes emission lines from highly ionised elements such as O, Ne, Mg, Si, S 
and
Fe. The well known distance of the SMC allows for good estimates of quantities 
such as physical size, mass, age and so on.

While the LMC SNRs have been relatively well studied, the remnants in the SMC 
have
not received much attention. For example, Hughes et al.~(\cite{hughes98})
conducted a systematic study of ASCA-SIS spectra of 7 SNRs in the LMC, while
Nishiuchi~(\cite{nishiuchi}) analysed 9 fainter remnants. They were able to
derive good estimates for the physical parameters of individual SNRs. In 
contrast,
detailed X-ray studies of the SMC have been limited to the youngest and
brightest SNR, IKT 22 (SNR 1E\,0102-72.3). This is mainly because SNRs in the 
SMC are fainter than their LMC counterparts.

To date a total of 16 SNRs have been identified from a number of extensive 
X-ray
surveys with the instruments aboard the \textit{Einstein}, \textit{ASCA} and
\textit{ROSAT} Observatories (e.g. Inoue et al.~\cite{inoue}; Wang \&
Wu~\cite{wang}; Haberl et al.~\cite{haberl}). The availability of the \chandra\
and \xmm\ Observatories now provides an opportunity to study these remnants 
with
far greater sensitivity and spatial/spectral resolution than before.

We present \xmm\ observations of 13 SNRs in the SMC. The major emphasis of this
work is on the spectral analysis of the CCD-resolution data obtained with the
European Photon Imaging Cameras (\epic) (Turner et al.~\cite{turner}; 
Str{\"u}der
et al.~\cite{struder}). In addition, we also present high resolution spectra of
IKT 22 and IKT 23 as measured by the Reflection Grating Spectrometers (\rgs) 
(den
Herder et al.~\cite{herder}).

\section{Observations and Reduction}

Our data are extracted from five pointings towards the SMC. The log of the
observations are given in Table 1. The pointings, as listed in Table 1, are
centred on the remnants IKT 5, IKT 18, IKT 23 and IKT 22 respectively.

The raw \epic\ data were initially processed with the \xmm\ Science Analysis
System (SAS) version 5.4. This involved the subtraction of hot, dead, or
flickering pixels, and the removal of events due to electronic noise. The
spectra were extracted using the SAS task EVSELECT. We used the standard
redistribution matrices available on the VILSPA site, while the ancillary 
matrices were 
created
using the SAS task ARFGEN. Background subtraction was done by selecting blank
sky regions within the same observation.

Mosaic images of the SMC fields were created using our own image and mosaicing 
software. The image extraction programme selects the photons with 
the appropriate grades and quality flags and saves the images in FITS file 
format. The latter is necessary in order to use the standard SAS EEXMAP 
programme
to make exposure maps for each extracted image. EEXMAP produces 
exposure maps normalized to the on axis exposure time, so each exposure map 
was multiplied with the appropriate \pn\ or \mos\ effective area by the 
mosaicing 
programme. The routines were called from a script that applied those programmes 
to 
each event-list and each channel range.

We also have high resolution spectra of IKT 22 and IKT 23 in addition to the 
CCD-resolution spectra. \rgs\ spectra of IKT 22 and IKT 23 were obtained with 
effective exposure times of 70 and 37 ks respectively. The \rgs\ data were 
processed with SAS version 5.4.

\begin{table}
\caption{\xmm\ \epic\ observation log.}
\label{tab:tab1}
\centerline{
\begin{tabular}{lcccccc} \hline
Rev Nr  & Obs ID      &  \multicolumn{2}{c}{Mode}& Filt. & 
\multicolumn{2}{c}{Exp. 
(ks)}  \\ 
         &              &   pn      &   MOS       &          &   pn     & MOS  
\\ \hline
156     & 0110000101   & EF & FF      & M   & 23& 27       \\
157     & 0110000201   & EF & FF      & M   & 16 & 20          \\
157     & 0110000301   & EF & FF      & M   & 31 & 35          \\
247     & 0135720601   & FF & LW      & T   & 16 & 33      \\
433     & 0135720901   & FF & LW      & T   & 14 & 13          \\ \hline
\multicolumn{7}{l}{ Abbreviations used} \\ 
EF & \multicolumn{5}{l}{Extended Full Frame} \\
FF & \multicolumn{5}{l}{Full Frame} \\
LW & \multicolumn{5}{l}{Large Window} \\
T & \multicolumn{5}{l}{Thin filter} \\
M & \multicolumn{5}{l}{Medium Filter}\\
\end{tabular}
}
\end{table}

\begin{figure*}
\resizebox{\hsize}{!}{\includegraphics{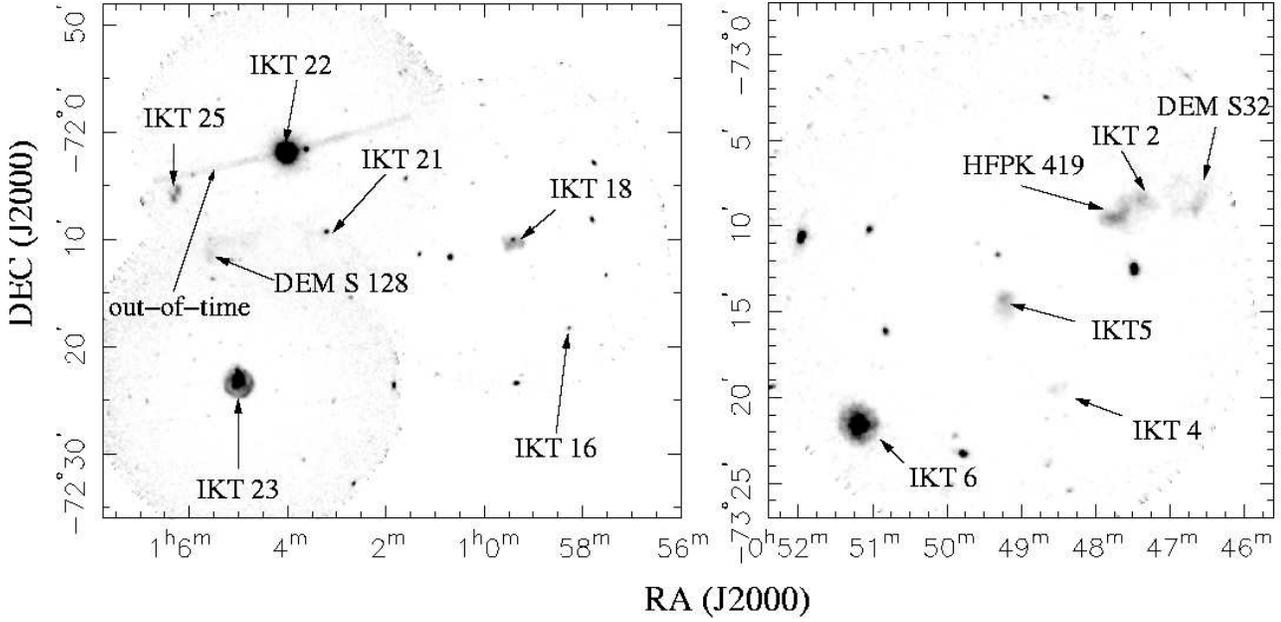}} 
\caption[]{Mosaic images of the SMC region. The left image is the combined 
observations of \xmm\ Rev's 157, 247 and 433, while the right panel is an image 
of Rev 156. Individual remnants are indicated. The out-of-time events from IKT 
22 can be seen .} 
\label{fig:field}  
\end{figure*} 

\begin{figure*}
\resizebox{\hsize}{!}{\includegraphics{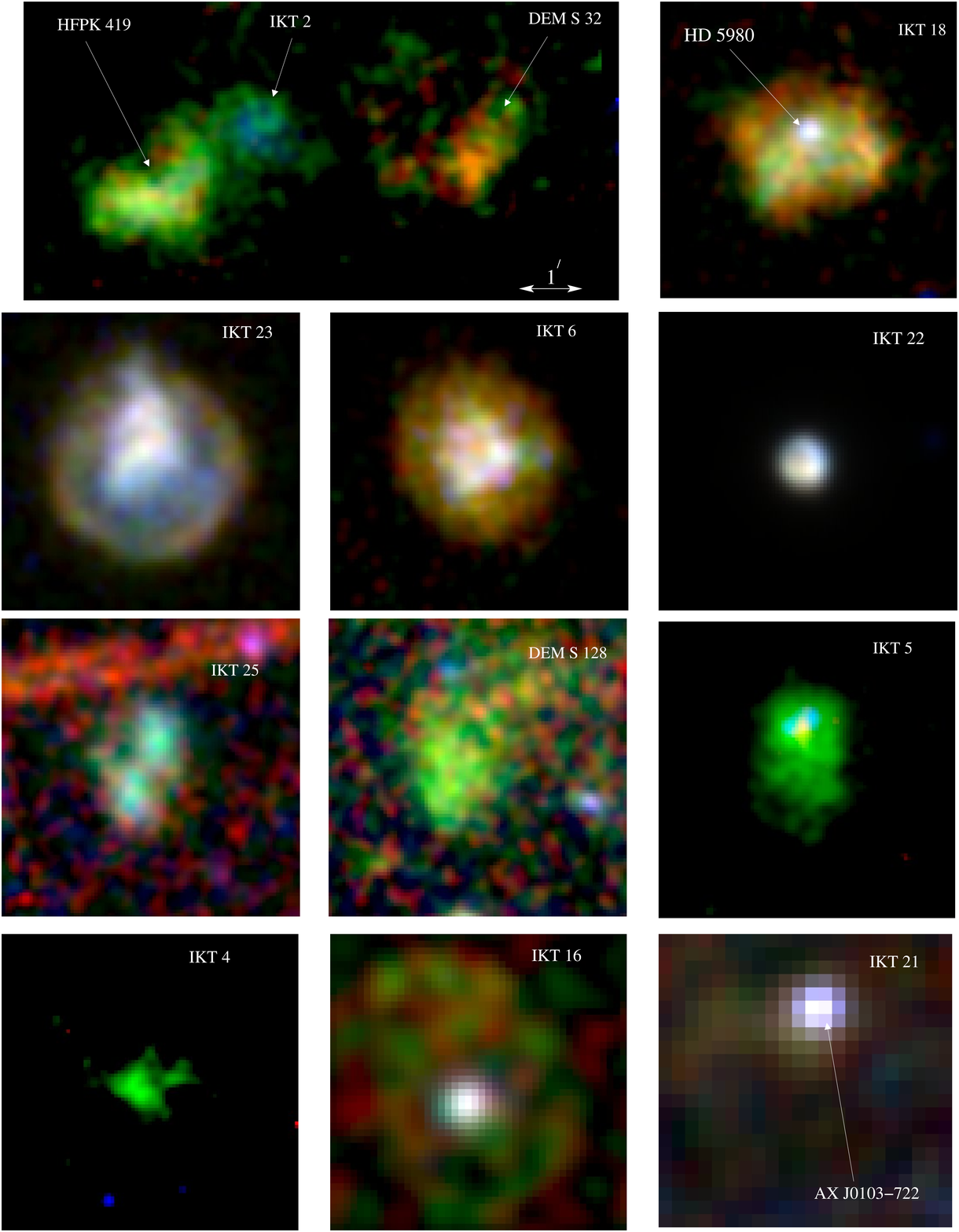}} 
\caption[]{Details of the remnants in our sample. Red, green and blue 
represents the 0.5--0.7~keV,0.7--1.0 and 1.0--3.0 keV energy bands. All the 
images are on the same plate scale.} 
\label{fig:detail}  
\end{figure*}

\section{Analysis and Results}

\subsection{Images}

In Fig.~\ref{fig:field} we present a mosaic of our observations. The left image 
is
a mosaic of \xmm\ revolutions 157, 247 and 433, while the right image shows the 
southwestern part of the SMC (Rev 156). We are able to identify 
13 SNRs in our field, these are indicated in Fig.~\ref{fig:field} 
and listed in Table~\ref{tab:tab1}. All the remnants are quite soft X-ray 
sources, with
the peak of their emission in the 0.5--1.0 keV band. A number of 
point sources, which generally display much harder spectra than the remnants, 
are also visible in 
the field-of-view. 
We present detailed colour coded maps of each remnant in Fig.~\ref{fig:detail}.
It is clear from Fig. 2 that the remnants exhibit a range of different 
morphological
features. The physical intepretation of these morphologies will be discussed in 
more
detail in Sect.~\ref{individual}.

\begin{table}
\caption{Log of the SNRs in our field.}
\label{tab:tab1}
\centerline{
\begin{tabular}{lcccc}
\hline
Object $^{*}$ &  SNR & Obs. ID    & RA         & DEC	     \\
       & Cat. &       &  (J2000)    & (J2000)      \\ \hline
DEM S\,32 & 0044-73.4 & 0110000101  & 00 46 39.1 & -73 08 39 \\
IKT 2  & 0045-73.4  & 0110000101   & 00 47 12.2 & -73 08 26 \\
HFPK 419&	    & 0110000101   & 00 47 40.6 & -73 09 30 \\
IKT 4  &  0046-73.5  & 0110000101   & 00 48 24.8 & -73 19 24 \\
IKT 5  &  0047-73.5  & 0110000101   & 00 49 06.9 & -73 14 05 \\
IKT 6  &  0049-73.6  & 0110000101   & 00 51 06.5 & -73 21 26 \\
IKT 16 &  0056-72.5  & 0110000201   & 00 58 16.4  & -72 18 05 \\
IKT 18 &  0057-72.2  & 0110000201   & 00 59 25.4 & -72 10 10 \\
IKT 21 &  0101-72.4  & 0123110201   & 01 03 12.8 & -72 08 59 \\
 IKT 22 &  0102-72.3  & 0123110201   & 01 04 02.0 & -72 01 48	\\
 IKT 23 &  0103-72.6  & 0110000301   & 01 05 03.5 & -72 22 56 \\
 DEM S\,128&  & 0123110201   & 01 05 23.2 & -72 09 26 \\
 IKT 25 &  0104-72.3  & 0123110201   & 01 06 14.3 & -72 05 18	 \\ \hline
\multicolumn{5}{l}{$^{*}$ Abbreviations used} \\ 
DEM S & \multicolumn{4}{l}{H$\alpha$ catalogue of emission 
nebulae (Davies et al.~\cite{davies}) } \\
IKT & \multicolumn{4}{l}{X-ray catalogue (Inoue et 
al.~\cite{inoue}) } \\
HFPK & \multicolumn{4}{l}{X-ray catalogue (Haberl et 
al.~\cite{haberl})} \\
\end{tabular}
}
\end{table}

\subsection{Spectral Analysis}

\begin{figure*}
\resizebox{\hsize}{!}{\includegraphics{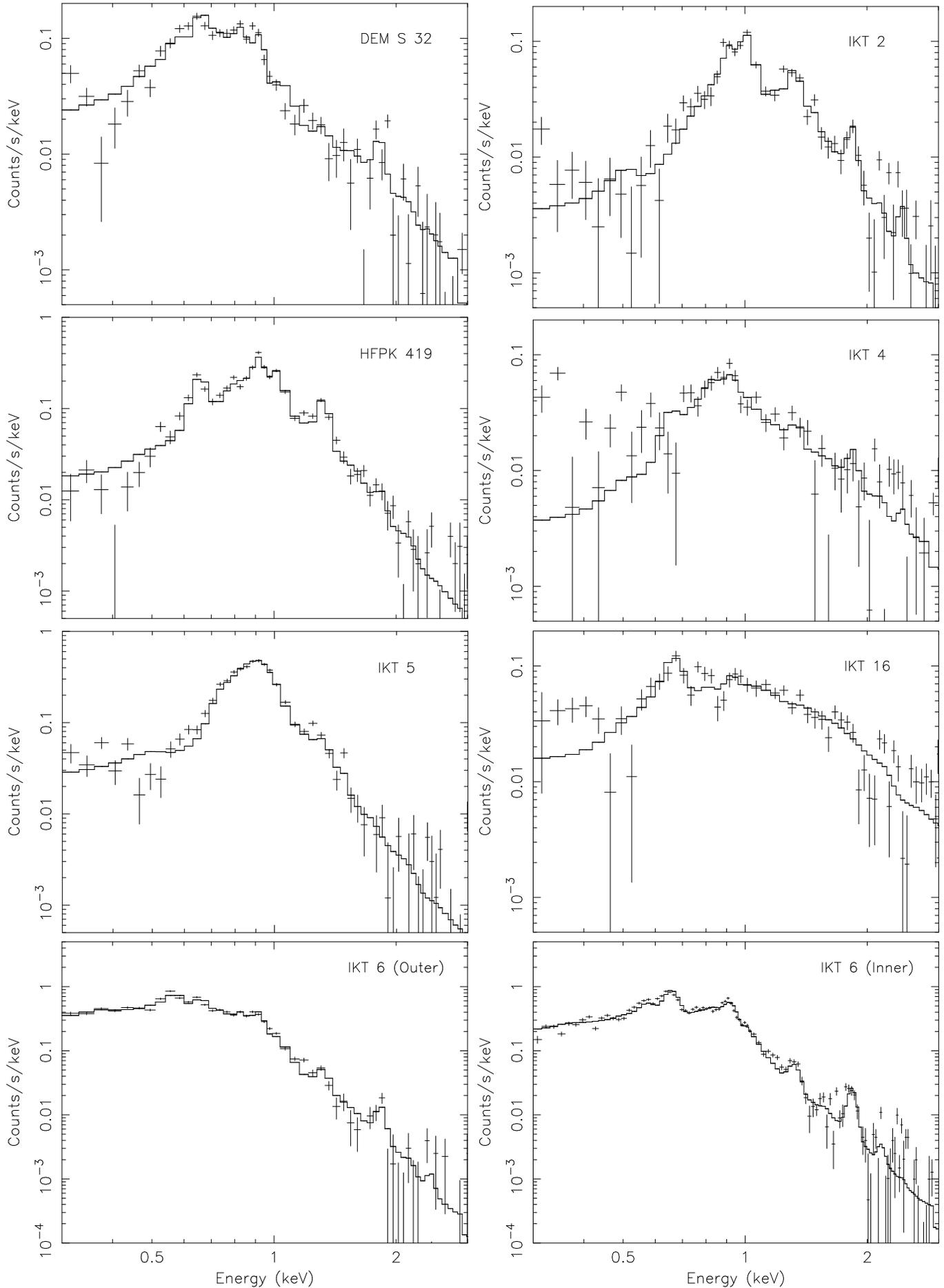}} 
\caption[]{Combined \epic\ (i.e. \pn+\mos) background-subtracted spectra of 
individual 
SNRs. The 
crosses represent the data points and error bars and the solid histogram 
represent the best fit models.} 
\label{fig:specplot1}  
\end{figure*} 

\begin{figure*}
\resizebox{\hsize}{!}{\includegraphics{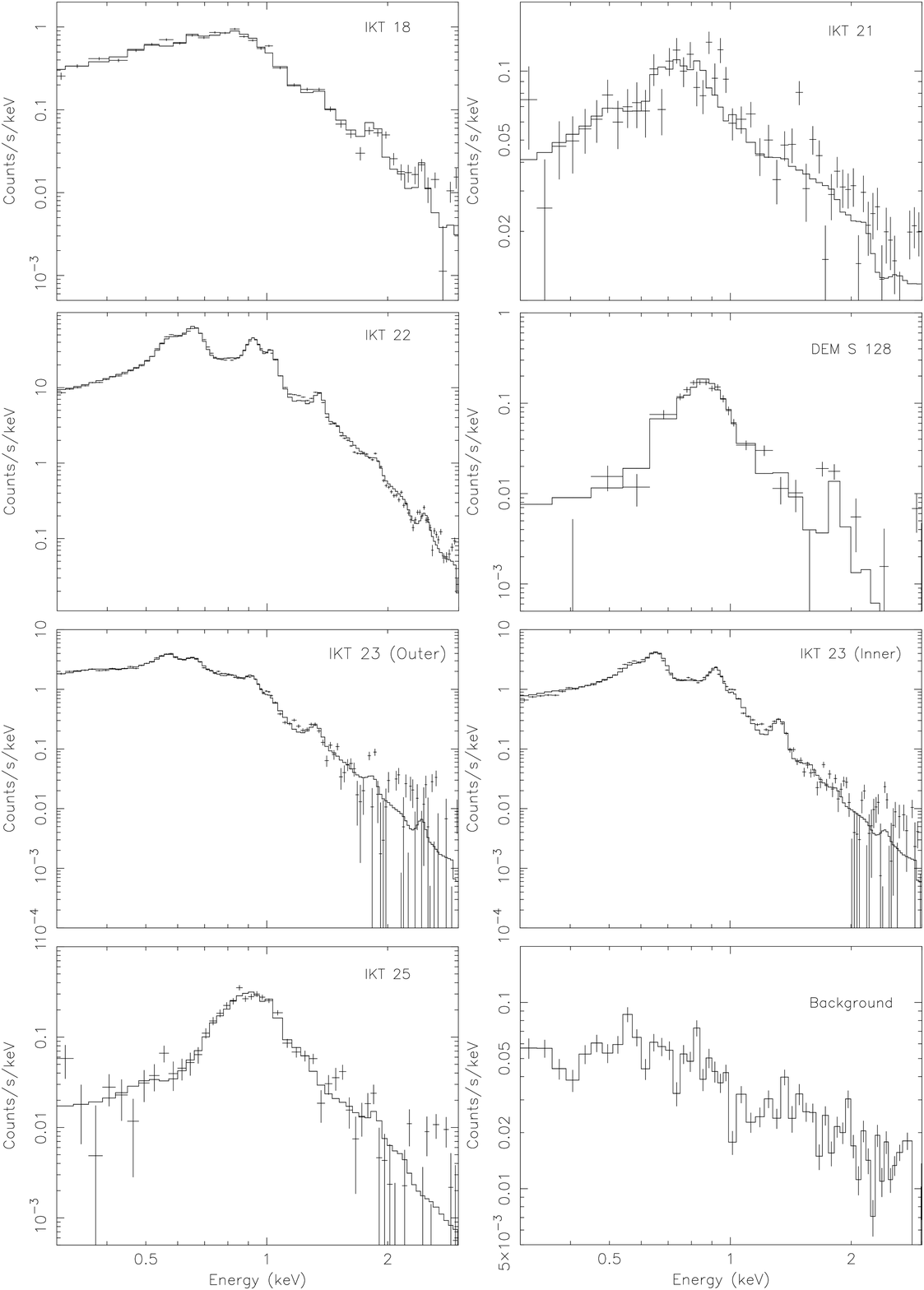}} 
\caption[]{Same as for Fig.~\ref{fig:specplot1}. The lower right panel 
represents a
typical background spectrum.} 
\label{fig:specplot2}  
\end{figure*} 

The spectral analysis was performed using the SRON SPEX package (Kaastra et
al.~\cite{kaastra}), which contains the MEKAL atomic database (Mewe et
al.~\cite{mewe}) for thermal emission. Also contained in SPEX are a
number of plasma models appropriate for SNR analysis.

We extracted spectra from the entire region of each SNR, while the background 
spectra were taken from nearby blank-sky regions. The spectra of individual 
remnants are shown in Figs.~\ref{fig:specplot1} and
\ref{fig:specplot2}. A typical background spectrum is also displayed in 
Fig.~\ref{fig:specplot2}. 
All our SNR spectra are thermal in nature and show emission from highly ionised 
atoms of 
O, Ne, Mg,
Si and Fe. 

We first attempted to fit all the spectra with a single temperature
non-equilibrium ionisation (NEI) model. This assumes that the plasma has been
instantly shock-heated to a temperature $T_{\rm e}$ some time ($t$) ago. The
NEI model does not contain any dynamical model for SNR evolution. In this model 
we
fit the volume emission measure ($n_{\rm es}n_{\rm Hs}V$), the electron
temperature ($kT_{\rm e}$), the ionisation parameter ($n_{\rm es}t$), the
elemental abundances and the column density $N_{\rm H}$ of absorbing gas along
the line of sight. $n_{\rm es}$ and $n_{\rm Hs}$ are the electron and hydrogen
density of the shocked gas, respectively and $V$ is the volume of emitting gas.
A distance of 60 kpc to the SMC is assumed. IKT 21 has a pulsar (AX J0103-722)
in the field. We have added a power-law component to account for the pulsar 
emission.

During the fitting procedure we initially kept the abundances fixed at the mean
SMC ISM value of 0.2 solar (Russel and Dopita~\cite{russel}). We started by
allowing only the normalisations of the plasma components to vary and 
subsequently we
also allowed the temperature, ionisation age, absorption and abundances (where
necessary) to vary. The spectra from \pn\ and \mos\ were fitted simultaneously.
The results of the fits are given in Table~\ref{results}. The abundances are
all relative to their solar values (Anders \& Grevese~\cite{anders}). We give
the unabsorbed luminosities in the 0.5--2.0 keV band.

We also estimated the electron density and radiating mass based on the best-fit
emission measure. Here it is provisionally assumed that the emitting gas is 
confined to shells of a typical thickness R/12. The radii were estimated from 
the X-ray images. The
mass estimates indicate that most of the sources are evolved remnants that
have swept-up a considerable amount of mass and should thus be emitting in 
their
Sedov phases. Moreover, a number of remnants could not be adequately described
by the simple NEI model, in particular, the spectra with higher statistics 
(viz,
IKT 6, IKT 22, IKT 23). These spectra appear to show an ionisation and/or
temperature distribution. We thus consider a Sedov model to be the most 
appropriate for fitting the data. This model is based on the
Sedov solution (Sedov~\cite{sedov}) which describes the self-similar expansion
of an adiabatic blast wave in a homogeneous medium.

We proceed to fit the IKT 6, IKT 18, IKT 22 and IKT 23 spectra with a Sedov 
model, based on unacceptable ${\chi}^{2}$'s and large swept-up mass, 
obtained from the NEI model fits. IKT 6 and IKT 23 have a shell and centrally 
filled morphology. We extract and fit only the shell part of the spectra of 
these two remnants with a Sedov model. Even though IKT 22 is clearly less 
evolved than IKT 6 and IKT 23 (for example), we do find that the Sedov model 
provides a much better fit to the data. Hayashi et al.~\cite{hayashi} also 
found that Sedov model fits to the {\textit ASCA} data of IKT 22 gave the most 
reliable results. We also fit, based on the high mass estimates for the NEI 
fits, the DEM S\,32, IKT 2, 
HFPK 419, and IKT 16 spectra with a Sedov model.

The free parameters of the Sedov model are the normalisation ($n_{\rm e}n_{\rm
H}R^{3}/d^{2}$), shock temperature ($T_{\rm s}$), ionisation parameter ($n_{\rm
e}t_{\rm i}$), elemental abundances and the column density $N_{\rm H}$ of
absorbing gas. Here $n_{\rm e}$ is the pre-shock electron density, $t_{\rm i}$
is the ionisation age of the remnant, $R$ is the SNR shock radius and $d$ is 
the
distance to the source. The fitting procedure for the Sedov models is the same 
as in the case of the NEI models. The fit results are summarised in 
Table~\ref{sedresults}.

From these fits, we can estimate several physical parameters for the SNRs in 
question.
If $n_{\rm e}$ is the electron
density, $n_{\rm H}$ is the hydrogen density, $n_{\rm m}$ is the total
number density of protons+neutrons (including those bound up in nuclei), then,
using the elemental abundances
and assuming a fully ionised plasma we can calculate the number of electrons
per hydrogen atom $r_{\rm e}=n_{\rm e}/n_{\rm H}$ and the effective number of
protons and neutrons
(baryon mass) per hydrogen atom $r_{\rm m}=n_{\rm m}/n_{\rm H}$. By adopting 
the 
values of the parameters mentioned above we estimate the electron density 
($n_{\rm e}$), the hydrogen density 
($n_{\rm H}$), the Sedov dynamical age 
($t_{\rm dyn}$), the effective ionisation age ($t_{\rm i}$), the total emitting 
mass ($M$) and initial explosion energy ($E_{0}$) by using:

\begin{equation}
n_{\rm e}=\sqrt{Nr_{\rm e}d^{2}/R^{3}}, \;{\mathrm m^{-3}} \label{eq1}
\end{equation} 
\begin{equation}
n_{\rm H}=\sqrt{Nd^{2}/(R^{3} r_{\rm e})}, \; {\mathrm m^{-3}} \label{eq2} 
\end{equation} 
\begin{equation}
t_{dyn}=1.3{\times}10^{-14}R/\sqrt{T}, \;{\mathrm yr}\label{eq3}
\end{equation}
\begin{equation}
t_{i}=3.17{\times}10^{-8}I_{\rm t}/n_{\rm e}, \;{\mathrm yr} \label{eq4}
\end{equation}
\begin{equation}
M=5{\times}10^{-31}m_{\rm p}r_{\rm m}n_{\rm H}V, \; {\mathrm M_{\sun}} 
\label{eq5}
\end{equation}
\begin{equation}
E_{0}=2.64{\times}10^{-15}TR^{3}n_{\rm H}, \; {\mathrm J} \label{eq6}
\end{equation}

Here $m_{\rm p}$ is the proton mass, $R$ is the shock radius, 
$V(=4/3{\pi}R^{3})$ is the total volume. We do not apply a volume filling 
factor to our estimates (eqs.~\ref{eq1}--\ref{eq6}) as it is already 
implicitely contained within the Sedov model. We assume that the remnants are 
spherically symmetric and therefore also do not apply a geometrical filling 
factor term to the volume estimate. The results are given in 
Table~\ref{tab:param}. The mass estimates based on the NEI fits compare
well to those obtained from the Sedov model fits.

We also fit the spectra from the inner regions of IKT 6 and 23. As a model we 
use one NEI component to account for emission from the central region. In 
addition to this, we added the same plasma component and parameters as derived 
from the fits to the outer region, only the normalisations from these 
components were allowed to vary. This method accounts for any projected 
fore/background shell emission. The fit results supplied in Table 
\ref{results}, are very similar between the two remnants. 

The best fit models generally provides good fits to the data, though a few 
discrepancies exist. Uncertainties in the modeling of the data include the 
calibration differences between the \pn\ and \mos\ and incompleteness of the 
atomic database. For example, the most prominent residual is the 
underestimation of the spectra at $\approx 1.2$\ keV. This is a known problem 
and is probably due to missing high excitation lines of \ion{Fe}{xvii-xix} in 
the plasma code (see Brickhouse et al.\cite{brickhouse}).

\begin{table*}
\caption{NEI model spectral fitting results. The fit errors ($1\sigma$) are 
given in 
brackets.
The abundances are all relative to their solar 
values (Anders \& Grevese~\cite{anders}).}
\centerline{
\begin{tabular}{lccccccc}
\hline
SNR        & Radius   & $N_{\rm H}$  & $n_{\rm e}n_{\rm H}V$ & $kT_{\rm e}$ & 
$n_{\rm e}t$ & $n_{\rm e}$ & Mass \\ 
           & (arcsec) &  ($10^{25}$ m$^{-2}$) & ($10^{64}$ m$^{3}$)& (keV) & 
($10^{16}$ m$^{-3}$s) & ($10^{6}$) m$^{-3}$ & M$_{\sun}$ 
\\ \hline  
DEM S\,32   & 68   &  2.4(0.7)   &  0.60(0.25)   &     1.51(0.48)   &  1.0(4.5) 
   &  
0.18  &  39  \\
IKT 2       & 33   &  3.9(1.5)   &  1.47(1.38)   &     0.77(0.59)   &  $>$20.0  
    & 
0.84  &  21   \\
HFPK 419    & 45   &  3.9(1.5)   &  1.21(1.1)	 &     0.62(0.39)   &  $>$13.0  
    & 
0.48  &  30   \\
IKT 4	    & 42   &  6.2(0.9)   &  0.14(0.03)   &     3.5(1.4)     &  2.1(0.2) 
   &  
0.18  &  9   \\
IKT 5       & 58   &  1.1(0.9)   &  0.47(0.23)   &     0.71(0.04)   &  $>$48.0  
    & 
0.20  &  27   \\
IKT 6       & 74   &  0.6(0.2)   &  5.89(0.8)	 &     0.54(0.03)   &  3.5(0.4 
)   &  
0.50  &  139 \\
IKT 16      & 100  &  0.6(0.5)   &  1.91(0.42)   &     1.6(0.4)     & 
0.37(0.07)   &  
0.17  &  104   \\
IKT 18      & 79   &  2.5(0.5)	 &  9.13(3.5)	 &     0.57(0.12)  &  2.7(+100)	
   &  
0.56  &  190 \\
IKT 21{*}   & 31   &  2.0(0.8)   &  0.48(0.31)   &     0.58(0.40)    &  
4.3(+100)   &  
0.53  &  11  \\
IKT 22      & 24   &  1.1(0.1)   &  73.2(5.6)	 &     0.38(0.01)   &  $>$600	
    & 
9.62  &  93   \\
IKT 23      & 99   &  0.41(0.12)  &   9.3(1.3)	 &     0.68(0.04)   &  2.5(0.4) 
   &  
0.41  &  270 \\
DEM S\,128  & 62   &  1.8(1.2)   &  0.17(0.1)	 &     0.61(0.1)    &  $>$200	
    & 
0.11  &  18   \\
IKT 25      & 55   &  4.8(1.7)   &  1.20(0.5)	 &     0.60(0.1)    &  38(20.0) 
   &  
0.36  &  40  \\ \hline
IKT6-inner & 30    & 0.8(0.1) 	&  0.13(0.06)	   &   0.89(0.16)  &  5.4(2.1) 
&0.29  &  
5 		 \\
IKT23-inner& 40    &1.1(0.1) 	&  0.22(0.08)	   &   0.92(0.12)  &  2.8(0.5) 
&0.25  &  
11	     \\ \hline
\end{tabular} 
}
\centerline{
\begin{tabular}{lccccccc} \hline
SNR       &	O       & Ne      & Mg     &   Si  & Fe   & L$_{\rm x}$(0.5-2 
keV)&${\chi}^{2}$/d.o.f 
\\ 
            &                &              &                 &               &          
    & ($10^{27}$ W) &  \\                 \hline
DEM S\,32   &    0.20(0.06)  &  0.23(0.08)  &   $<$0.10)      &  0.29(0.21)   &  
0.08(0.05)  & 19  & 190/151  \\
IKT 2       &    $<$0.10      &  0.81(0.42)  &   0.72(0.35)  &  0.41(0.13)   &  
0.13(0.09)  & 14  & 152/155  \\
HFPK 419    &    1.21(0.9)   &  1.91(1.5)   &   1.64(1.4)   &  0.24(0.2)    &  
0.39(0.34)  & 42  & 192/155  \\
IKT 4	    &    0.2(f)      &  0.2(f )     &   0.2(f)      &  0.2(f)	    &  
0.2(f)	 
  & 18  & 205/155  \\
IKT 5       &    $<$0.1	     &  0.92(0.65)  &   0.93(0.45)  &  $<$0.1	    &  
0.91(0.35)  & 9.6 & 340/320  \\
IKT 6       &    0.11(0.01)  &  0.24(0.03)  &   0.17(0.04)  &  0.42(0.2)    &  
0.09(0.01)  & 66  & 610/320  \\
IKT 16      &    0.20(f)     &  0.20(f)     &   0.20(f)     &  0.20(f)      &  
0.20(f)   
  & 61  & 162/109  \\
IKT 18      &    0.06(0.02)  &  0.09(0.04)  &   0.07(0.03)  &  0.25(0.1)    &  
0.07(0.1)	 
  & 62  & 333/153  \\
IKT 21      &    0.20(f)     &  0.20(f)     &   0.20(f)     &  0.20(f)      &  
0.20(f)   
  & 8.0 & 416/361  \\
IKT 22      &    2.03(0.14)  &  2.46(0.24)  &   1.50(0.06)  &  0.65(0.09)   &  
0.10(0.01)  & 150 & 3043/212  \\
IKT 23      &    0.14()      &  0.28()      &   0.21()      &  0.00()	    &  
0.07()	 
  & 150 & 1103/320  \\
DEM S\,128  &    0.00()      &  0.00()      &   2.60(1.8)   &  2.43(1.6)    &  
2.34(1.1) 
  & 7.8 & 125/115  \\
IKT 25      &    $<$0.1	     &  3.01(1.2)   &   $<$0.1	    &  $<$0.1	    &  
1.53(0.4) 
  &  45 &  139/156  \\ \hline
 \hline 
IKT6-inner  &   2.0(0.6)      &   2.3(0.8)  &   1.6(0.9)  &  2.7(1.2)  &  
0.87(0.31)  &  
17 &180/155            \\
IKT23-inner &   2.1(0.5)      &   4.7(1.2)  &   3.3(1.3)  &  0.2(f)     &  
0.55(0.15)  & 
37  & 410/320          \\ \hline
\multicolumn{8}{l}{* plus power-law component, norm.$\sim 2.2{\times}10^{43}$ 
phs$^{-1}$keV$^{-1}$ and $\gamma \sim 0.9$}\\
\multicolumn{7}{l}{$^{+}$f - fixed}
\end{tabular}
}
\label{results}
\end{table*}

\begin{table*}
\caption{Sedov model spectral fitting results. The fit errors ($1\sigma$) are 
given in 
brackets. The 
abundances are all relative to their solar 
values (Anders \& Grevese~\cite{anders}).}
\label{tab:}
\centerline{
\begin{tabular}{lccccc}
\hline
SNR        & $N_{\rm H}$  & $n_{\rm e}n_{\rm H}R^{3}/d^{2}$ & $kT_{\rm e}$ & 
$n_{\rm e}t$ & L$_{\rm x}(0.5-2 keV)$ \\ 
            &  ($10^{25}$ m$^{-2}$) & ($10^{20}$ m$^{3}$)& (keV) & 
($10^{16}$ m$^{-3}$s) & ($10^{27}$ W)\\ \hline  
 DEM S\,32  &1.9(0.6)  & 1.70(0.7)   & 1.29(0.46)  & 2.5(1.1)  & 16 \\
 IKT 2      &5.1(1.2)  & 16.3(9.3)   & 0.39(0.05)  & $>$220  & 20 \\
 HFPK 419   &4.4(0.9) & 12.7(5.2)  & 0.28(0.03)  & $>$360  & 57 \\
 IKT 6      &0.7(0.25)  & 22.5(7.5)   & 0.27(0.05)  & 11.1(0.5)  & 39 \\
 IKT 16     &5.1(0.6)  & 4.32(1.8)   & 1.76(0.65)  & 0.51(0.1) & 39 \\
 IKT 18     &1.3(0.25)  & 15.4(5.1)   & 0.51(0.06)  & 20.0(10.1) & 39 \\
 IKT 22     &0.60(0.06)  &  99.3(11)  &  0.78(0.08)  & 9.0(+1.7)  & 
$1.3{\times}10^{3}$ \\
 IKT 23     &0.5(0.1)  & 51.9(5.8)   & 0.26(0.03 )  & 20.0(0.4) & 85 \\ \hline
\end{tabular} 
}
\centerline{
\begin{tabular}{lcccccc} \hline
SNR       &	O       & Ne      & Mg     &   Si  & Fe   & ${\chi}^{2}$/d.o.f 
\\ \hline
 DEM S\,32 & 0.28(0.15)  & 0.17(0.14) &  0.19(0.14) &  0.45(0.26)   &0.23(0.1)  
  &     
164/149  \\
 IKT 2     & $<$0.08  & 0.89(0.39) &  0.51(0.23) &  0.33(0.17)   & $<$0.1    &     
152/149  \\
 HFPK 419  & 1.25(0.45)  & 2.19(0.74) &  1.65(0.54) &  0.13(0.1)   &0.25(0.1)   
 &     
193/150  \\
IKT6-outer  &   0.13(0.03)    &   0.28(0.08)&   0.3(0.12) &  0.4(0.3)  & 
0.15(0.04)     &     223/155  \\
 IKT 16    & 0.2(f)  & 0.2(f) &  0.2(f) &  0.2(f)   &0.2(f)    &     199/165  
\\
 IKT 18    & 0.14(0.07)  & 0.11(0.07) &  0.22(0.09) &  0.44(0.13)   &0.17(0.03) 
   &     
253/153  \\
IKT 22    & 0.87(0.05) &  1.99(0.10)  & 1.32(0.09)   &0.21(0.03)  & 0.29(0.02)  
&  841/211 
\\
IKT23-outer &   0.17(0.02)    &   0.28(0.04)&   0.35(0.07)&  0.36(0.15)&   
0.11(0.02)    &     460/320  \\ \hline
\end{tabular}   
}
\label{sedresults}
\end{table*}

\begin{figure}
\resizebox{\hsize}{!}{\includegraphics[angle=-90]{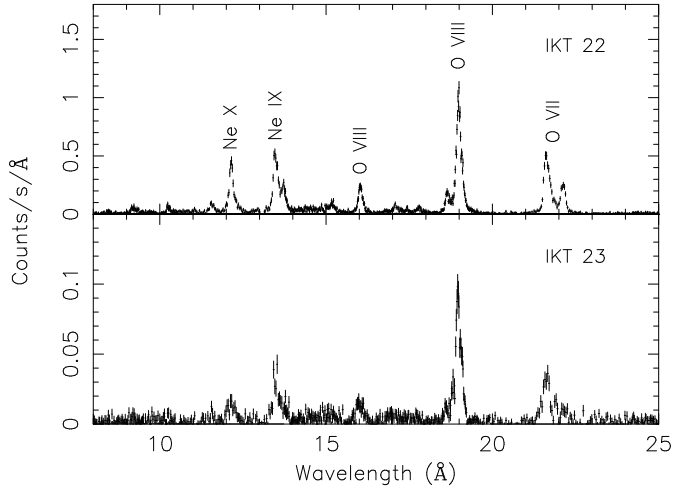}} 
\caption[]{\rgs\ spectrum of IKT 22 (top) and IKT 23 (bottom). The most 
prominent emission lines are labeled.} 
\label{fig:rgs}  
\end{figure}

\begin{table}[t!]
\caption{Physical parameters derived from the best fit Sedov model results.}
\label{tab:param}
\centerline{
\begin{tabular}{lcccccc}
\hline
SNR      &  $n_{\rm e}$    &  $n_{\rm H}$     & $t_{\rm i}$   & $t_{\rm dyn}$   
  & Mass 
   &  E$_{0}$  \\ 
         &\multicolumn{2}{c}{($10^{6}$ m$^{-3}$)} & ($10^{3}$ yrs) & ($10^{3}$ 
yrs) & 
(M$_{\sun}$) &  ($10^{44}$ J) \\ \hline 
 DEM S\,32  &  0.05 &0.05 &	 15    &      6   &   43      &    0.9     \\
 IKT 2      &  0.45 &0.44 &   $>$70    &      5.4 &   47      &    0.3     \\
 HFPK 419   &  0.26 &0.25 &   $>$100   &     8.5   &  64      &    0.6     \\
 IKT 6      &  0.21 &0.17 &	18    &      14  &   180      &    0.8     \\
 IKT 16     &  0.05 &0.04 &	3.5   &       7.5  &   124      &    3.4     \\
 IKT 18     &  0.13 &0.11 &	23    &      11  &   196      &    1.2     \\
 IKT 22     &  2.03 &1.68 &    1.4    &      2.7 &   84       &    0.8  \\
 IKT 23     &  0.19 &0.16 &	30    &      19  &   424      &    1.8     \\
 \hline
\end{tabular}
\label{tab:param}
}
\end{table}

\section{Discussion } 

\subsection{Progenitor Types{\label{individual}}}

\subsubsection{Core--Collapse candidates}

IKT 2 and HFPK419 are located in the large emission nebuale N19. These 
remnants
have rather irregular and complicated X-ray morphologies. For example, it is
not clear whether IKT 2 and HFPK 419 are indeed two seperate remnants or part 
of
a larger SNR. However, the spectra and derived abundances from these remnants 
are rather different. HFPK 419 shows enhanced O emission, while IKT 2 has 
virtually no
discernible O emission. Also, IKT 2 has more enhanced Si emission. These
differences presumably indicate that they indeed are separate objects. IKT 2 
and
HFPK 419 both show high metal abundances, which indicate that much of the X-ray 
emission is from ejecta enriched material. Their abundance profiles show a 
similar pattern in that the alpha-rich products O(except for IKT 2), Ne and Mg 
are higher than Fe. Enhanced O, Ne and Mg abundances usually indicates a 
core-collapse origin (e.g. Nomoto et al.~\cite{nomoto}; Woosley \& 
Weaver~\cite{ww95}). We thus suggest that these remnants are the results of 
core-collapse SNe. This is also in agreement with their surroundings which 
suggest at least modest starburst activity (Dickel et al. \cite{dickel}).

DEM S\,32 is also in the vicinity of N19. It has a shell-like 
structure with enhanced emission towards the southwestern limb. The abundance 
yields from DEM S\,32 are lower than for IKT 2 and HFPK 419 and no clear 
distinction in progenitor type can be infered from these yields. The starburst 
vincinity, however, suggest that DEM S\,32 possibly also originates from a 
core-collapse progenitor.

IKT 6 and IKT 23 are remarkably similar remnants, both morphologically and
spectrally (Figs.~\ref{fig:specplot1} and \ref{fig:specplot2} ). IKT 23 has a
clear shell and a centrally filled morphology. The colours in
Fig.~\ref{fig:detail} indicate that the outer region is softer than the inner
region. IKT 6 exhibits a similar morphology, although it is not as clearly
resolved. IKT 6 is located towards the edge of the detector where the effective
area and spatial resolution is lower. The other remnant in our sample with a
clear shell structure is IKT 22. This remnant is quite striking since it 
is orders of magnitude more luminous than the other SNRs in the field.

The spectra extracted from the two regions (i.e. outer and inner) of IKT 6 and
IKT 23 (see Figs.~\ref{fig:specplot1} and \ref{fig:specplot2}) are distinctly
different. However, the spectra are remarkably similar between the two
remnants. The spectral fitting results given in Tables~\ref{results} and
\ref{sedresults} also yield similar parameters for the two remnants. Although 
not clearly visible in the \xmm\ data (because of the lower spatial resolution), 
the \chandra\ data do reveal that IKT 22 has a blastwave and reverse shock 
structure (Hughes et al.~\cite{hughes00}). The spectra from these two regions 
are also distinctly different. 

The abundance derived from the fits to the
outer regions are consistent with the SMC ISM abundances, while the inner
regions reveal much higher abundance values. We propose, based on the abundance
profiles, that the inner region represents reverse shock heated ejecta material
while the outer region represents a blastwave moving through the ISM/CSM. The 
fit results also show that the inner (ejecta-rich) regions are hotter ($kT \sim 
0.9$ keV) than the outer regions ($kT \sim 0.27$ keV). This is expected from 
models such as proposed by Truelove and McKee~\cite{truelove}, which predict 
that the 
blastwave in more evolved remnants eventually attains a lower velocity than the 
reverse shock. 

The spectra extracted from the inner regions of IKT 6 and IKT 23 are also
similar to the IKT 22 spectrum. All three remnants show enhanced O, Ne and Mg 
abundances with respect to Fe. The abundance values of these elements (w.r.t. 
solar) are also an order of
magnitude larger than their SMC ISM values. We also obtained \rgs\ spectra of
IKT 22 and IKT 23. The spectra were extracted from the entire remnants. The
high-resolution \rgs\ spectra of IKT 22 and IKT 23, displayed in
Fig.~\ref{fig:rgs}, again bear striking similarities and reveal prominent
emission lines of O and Ne species. The most obvious difference between the two
spectra is in the count rate. The flux in the emission lines of IKT 22 are an
order of magnitude larger than compared to IKT 23. Much of this flux difference
can be attributed to the differences in abundance values. This is because the 
\rgs\ (and \epic) spectrum of IKT 22 is dominated by emission from ejecta-rich 
material, while the IKT 23 spectrum is dominated by emission from the swept-up 
ISM material, which has an average abundance of $\sim 0.2$ solar. Fits to the 
\epic\ data extracted from the entire remnant also show that the abundances for 
IKT 22 are an order of magnitude larger than compared to IKT 23. A more 
detailed \rgs\ analysis of IKT 22 is provided by Rasmussen et 
al.~(\cite{andy_1e0102}).

The similarities between IKT 6, IKT 22 and IKT 23 suggests that these three
remnants have common progenitors. IKT 22 is a well studied remnant that has
been classified as being the result of a type II SN. IKT 6 and IKT 23 should
thus also be the remnants of type II SNe. Both the derived abundances and
masses for the inner regions of IKT 6 and IKT 23 are consistent with this
interpretation. These remnants represent very different stages of SNR
evolution. IKT 22 is a $\sim$2\,000 yr old remnant (Blair et al.~\cite{blair}) 
while IKT 6 an IKT 23 are much
more evolved (14\,000--20\,000 yr). We propose that IKT 6 and IKT 23 are
evolved versions of IKT 22 and studying these remnants in more detail presents 
a
unique opportunity to probe the evolution of oxygen-rich remnants in the SMC.
It is also interesting to note that the ejecta are still visible in remnants as
evolved as IKT 6 and IKT 23.

IKT 16 is one of the weaker X-ray sources in our sample. The X-ray morphology 
appears consistent with the radio and H$\alpha$ classification of it being 
a shell like SNR (Mathewson et al.~\cite{mathewson84}). A harder emission spot 
can be seen in the centre of the X-ray image. We attempted to extract a separate
spectrum from this region, but the data is of insufficient quality to speculate
on the X-ray nature of this region. A longer exposure observation is needed to
investigate whether this source is thermal or non-thermal in nature.

Also interesting is the high explosion energy (${\sim}3.3{\times}10^{44}$ W) 
and
low ionisation age ($t_{\rm i}{\sim}3\,500$ yr) associated with IKT 16. The fit
parameters are not very well constrained due to the low count rate so the
discrepancies might originate from statistical limitations. However, in their
analysis of the LMC remnants, Hughes et al.~\cite{hughes98} found a connection
between low ionisation age (as compared to the dynamical age) and high explosion
energy. They explain this connection as SNRs which exploded within pre-existing
low-density cavities in the ISM. The remnant expands rapidly to the cavity
wall, where it then encounters denser gas and begins to emit X-rays. This
results in a lower dynamical age and explosion energy than that inferred from
the Sedov model. A similar scenario could thus be applicable to IKT 16. If
this is the case, then IKT 16 would be the result of the core-collapse of a
massive star whose stellar wind has caused a low density circumstellar
cavity.

There is no clear X-ray emission associated with the optical emission from IKT
21, however, our X-ray image (Fig.~\ref{fig:detail}) does show some faint,
diffuse emission extending to the east of the Be-pulsar binary system AX
J0103-722 (Israel et al.~\cite{israel}). The X-ray emission from the vicinity
of the optical remnant is, however, dominated by the puslar system. 

We extracted our spectrum from the region containing the optical emission, 
which includes the
pulsar system. Although not visible in Fig.~\ref{fig:specplot2}, the spectrum 
clearly extends above 3 keV. Our
model for IKT 21 thus consists of a NEI component for thermal emission from the
SNR and a power-law component to account for the pulsar emission. The power-law
component normalisation and photon index were initialy fixed to values obtained
from Hughes and Smith~\cite{hughes94} and were subsequently allowed to vary.
The best fit model gives a power-law component normalisation of $\sim
2.2{\times}10^{43}$ phs$^{-1}$keV$^{-1}$, a photon index of $\sim 0.9$ and a
luminosity of $\sim5.5{\times}10^{27}$ W(0.5--2.0 keV band). Our results for 
the thermal emission indicate a temperature of $kT_{\rm e}{\sim}0.58{\pm}0.40$ 
and a 
luminosity of $\sim8{\times}10^{27}$ W (0.5--2.0 keV band). These results 
compare well with estimates obtained by Hughes and Smith~\cite{hughes94}. It is 
not clear whether the pulsar AX J0103-722 is associated with the SNR as it is 
not near the centre of the remnant. An association would mean that IKT 21 would 
be the remnant of a massive-star SN (Hughes and Smith~\cite{hughes94}).

\subsubsection{Type Ia candidates}

The images in Fig.~\ref{fig:detail} show that IKT 4, IKT 5, DEM S\,128 and IKT
25 have similar diffuse X-ray morphologies with no clear shell like structure.
The emission peaks in the 0.7--1.0 keV band. This indicates that the flux of
these sources are dominated by emission from Fe-L transitions. These four
remnants also have similar spectral features. They show a broad
spectral hump around the Fe-L complex ($\sim 0.9$ keV). The excess emission
around $\sim 0.9$ keV and the high Fe abundance in these remnants strongly 
suggests that these remnants are the result of type Ia supernovae.

The X-ray sizes of IKT 4, IKT 5, DEM S\,128 and IKT 25 are much smaller than
their optical diameters (see Mathewson et al.~\cite{mathewson83} and Mathewson
etal.~\cite{mathewson84}). The large optical diameters imply that they are
evolved remnants. Mathewson et al.~(\cite{mathewson84}) also attributes the
high \ion{S}{ii} emission detected in IKT 5 and IKT 25 to the onset of 
radiative
cooling due to the evolved nature of these remnants. A possible interpretation
is that these remnants are much more evolved versions of SNR DEM L\,71 in the
LMC (Hughes et al.~\cite{hughes03};van der Heyden et al.~\cite{vanderheyden}).
Here we see a bright ISM dominated shell with fainter Fe-rich ejecta material 
in
the centre. In this scenario, IKT 4, IKT 5, DEM S\,128 and IKT 25 would have 
evolved to such a degree that the shells have eventually become too faint or 
too cool to be seen in X-rays and we only see the hotter Fe-rich ejecta 
remains. The morphologies and temperature structure of IKT 6 and IKT 23 also 
add credence to such a scenario.

The mass estimates given in Table~\ref{results} was made on the assumption that
the X-ray emission is from shock heated ISM material. However, if the emission
is from the ejecta remains then the metals (and not H) would be the main source
of the X-ray emission. This will alter our mass estimates. The mass, however,
depends on the amount of H mixed into the ejecta during its evolution (see
Hughes et al. \cite{hughes03} for a complete explanation). If we assume that a
comparable amount (in mass) of hydrogen has been mixed into the metal rich
ejecta then this would give Fe mass estimates of 0.4, 0.8, and 0.67 M$_{\sun}$
for IKT 5, IKT25 and DEM S\,128 respectively. These estimates are within the Fe
mass range given by Nomoto et al.~(\cite{nomoto}) for various type Ia SNe
models. We do not make mass estimates for IKT 4 since the quality of the data
does not allow for accurate abundance determination.

\subsubsection{The special case of IKT 18}

IKT 18 has a rectangular shape. The emission is diffuse with no obvious limb
brightening. The \chandra\ data (Naz{\'e} et al. \cite{naze}) do reveal a few
bright or dark arcs, but apart from these, the brightness is rather uniform.
The luminous blue variable (LBV) HD 5980 can also be seen in the centre of IKT
18. The spatial coincidence of IKT 18 with the peculiar binary system HD 5980
suggests an association with this system. The possible association was recently
investigated by Naz{\'e} et al.~(\cite{naze}). They drew attention to
similarities between IKT 18 and the Carina nebula. However, they concluded that
based on the non-thermal radio emission (Ye et al.~\cite{ye}) and a high
velocity expansion (Chu \& Kennicutt~\cite{chu}), IKT 18 should be regarded as 
a
SNR with HD 5980 located behind the remnant. The low SMC-like abundances and 
high mass (196 M$_{\sun}$) and age estimate (11\, 000 yr) derived from our fits 
are also consistent with the picture of an evolved SNR which has swept-up ISM 
material. The low abundances, however, do not allow for a progenitor typing. 
More \xmm\ data will be available in the near future, which will allow for a 
more detailed assessment of the nature of this object.

\subsection{SMC Abundances}

The abundance values for the swept-up matter dominated SNRs
(Table~\ref{sedresults}) show some spread among the remnants and also
between different elemental species. The general trend is that the larger
remnants have lower abundance values while the smaller ones have larger
values. The interpretation is that the older remnants have swept-up so much ISM
material that the abundances are approaching the ISM values. The abundance
yields from the older remnants with larger swept-up masses could thus be used 
to
probe the ISM abundance distribution.

\begin{figure}
\resizebox{\hsize}{!}{\includegraphics[angle=-90]{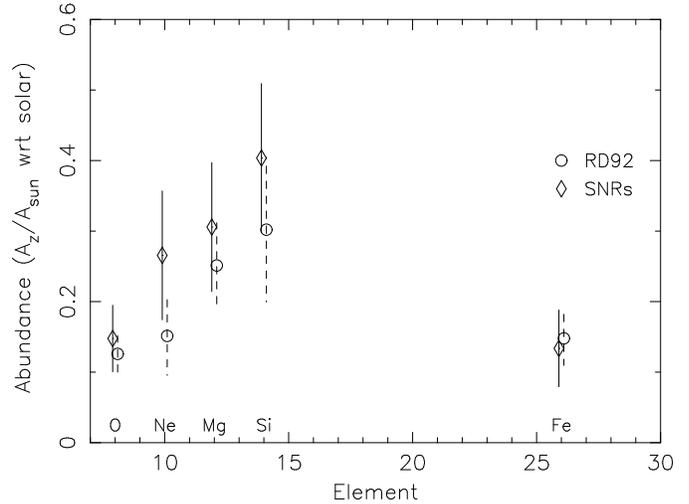}}
%\resizebox{\hsize}{!}{\includegraphics[angle=-90]{ave2.ps}} 
\caption[]{Average SMC abundances. The average abundance values obtained from 
fits to IKT 6 IKT 18 and IKT 23 are represented by diamonds. Results from 
Russel \& 
Dopita ~(\cite{russel}) are indicated with circles.} 
\label{fig:ism}  
\end{figure}

We use the abundances derived from the spectral fits to IKT 6 (outer region), 
IKT 18 and IKT 23 (outer region) to determine the gas-phase abundances of the 
SMC. These remnants are selected
because of their high (swept-up) masses. We average, for each
elemental species, the fitted abudances from the three remnants. In
Fig.~\ref{fig:ism} we plot the error weighted average abundances versus
elemental species. The uncertainties represent the rms-scatter among the 
remnants. The derived abundances are also compared to the average SMC abundance 
values presented by Russel \& Dopita~(\cite{russel}). The work of Russel \&
Dopita~(\cite{russel}) is based on the spectral analysis of F supergiants for
heavier elements (Z$>12$) and \ion{H}{ii} regions and SNRs for lower Z 
elements.
However, they only have 1 SMC SNR in their studies. Our average abundance
values for the low-Z elements appear slightly higher than those obtained by 
Russel \& Dopita~(\cite{russel}), but are in good agreement within the 
uncertainties.

\subsection{Physical parameters}

The ionisation ages given in Table~\ref{tab:param} are generally consistent 
with the dynamical age within a factor 2, the exceptions being IKT 2 and HFPK 
149. The factor of 2 difference can be easily explained invoking fragmentary 
shells or clumping of the X-ray emitting gas. This would increase the electron 
density and thus reduce the ionisation age estimates. The possibility of 
fragmentary shells or clumping means that the initial explosion energy $E_{0}$ 
and mass estimates should be regarded as lower and upper bounds respectively. 

In the case of IKT 2 and HFPK 149 the plasma has reached ionisation 
equilibrium. In this regime the ionisation parameter becomes insensitive as an 
age indicator since the spectrum changes very slowly with increasing ionisation 
parameter.

The mean SMC ISM hydrogen density derived from the Sedov model fits is $\sim
0.32{\times}10^{6}$ m$^{-3}$. This is nearly an order of magnitude lower than
the mean density of $\sim 1.8{\times}10^{6}$ m$^{-3}$ derived from Sedov fits 
to
a sample of LMC remnants by Hughes et al.~\cite{hughes98}. The lower SMC
abundance easily explains the lower SNR luminosities compared to their LMC
counterparts, as the X-ray emission scales with the square of the densities.
The lower densities also have implications for the dynamical evolution of SNRs 
in
the SMC. Falle et al.~(\cite{falle}) predicts that the onset of radiative
cooling occurs at a time $t_{\rm cool}=
3.55{\times}10^{7}E_{44}^{0.24}n_{H}^{-0.52}$ yr, where $E_{44}$ is the
explosion energy in $10^{44}$ W and $n_{H}$ is in m$^{-3}$. The onset of
radiative cooling will be governed by the hydrogen density, since the explosion
energies derived from the SMS and LMC samples are similar. The lower hydrogen
density means that the SMC remnants takes $\sim 2.5$ times longer to reach the
radiative cooling stage, compared to the LMC remnants. The SMC remnants should
thus be radiating X-rays for a longer time than in the LMC. This point is
illustrated by the fact that oldest remnant in our sample is $\sim$20\,000 yr
while the oldest remnant in the LMC sample analysed by Hughes et
al.~\cite{hughes98} is 10\,000 yr.

\subsection{SN rates}

We identified 8 core-collapse (II+Ib/c) and 4 type Ia remnants in 
Sect.~\ref{individual} and make simple SNe rate estimates, since the statistics 
do not allow for more involved calculations. By dividing the number of SNRs by 
the age of the oldest remnant we get a core-collapse SNe rate of one per 
2\,400$\pm$800 yr. The error accounts for the statistical uncertainty and
incompleteness of our sample. This is much lower than an estimate by Filipovic
et al.~(\cite{filipovic}) who give a birth-rate of one SNR per 350 yr. Their
result is based on a relation between the SNR age and radio flux. Our result is
more in line with a rate of one in every 1100--2500 inferred from van den Berg
\& Tammann~(\cite{berg}), which is derived from the star formation rate 
inferred from the total H$\alpha$ luminosity of the SMC. The derived rate is 
also 
consistent with our age estimate ($\sim 2\,700$ yr) of the youngest SMC SNR 
(viz. IKT 22). The number of type Ia SNRs is approximately 0.5 that of 
core-collapse, which implies a frequency of one in every 4\,800$\pm$1\,600 yr.

\begin{acknowledgements}
We thank Jacco Vink for the use of and help with his image processing software 
and for valuable discussions. The results presented are based on 
observations obtained with XMM-Newton, 
an ESA science mission with instruments and contributions directly funded by 
ESA Member States and the USA (NASA). SRON is supported financially by NWO,
the Netherlands Organisation for Scientific Research.
\end{acknowledgements}

\end{document}